\documentclass[11pt,a4paper,onecolumn,notitlepage]{article}
\usepackage[affil-sl]{authblk}
\usepackage[latin1]{inputenc}
\usepackage{amsmath}
\usepackage{amsfonts}
\usepackage{amssymb}
\usepackage{graphicx}
\usepackage{pdflscape}
\usepackage{tikz}
\usepackage{pgfplots}
\usepgfplotslibrary{statistics}
\usetikzlibrary{calc}
\usepackage{cisco}

\usepackage{fancyhdr}
\usepackage{hyperref}

\title{Modeling and Analysis of SDN Control Applications using Vector Spaces}
\author[1]{Mohamed Aslan\thanks{maslan@sce.carleton.ca}}
\author[2]{Ashraf Matrawy\thanks{ashraf.matrawy@carleton.ca}}
\affil[1]{Department of Systems and Computer Engineering, Carleton University. Ottawa, ON. Canada}
\affil[2]{School of Information Technology, Carleton University. Ottawa, ON. Canada}

\providecommand{\keywords}[1]{\textbf{\textit{Index terms---}} #1}

\newtheorem{definition}{Definition}
\newtheorem{axiom}{Axiom}

\newtheorem{corollary}{Corollary}

\begin{document}
\maketitle

\begin{abstract}
	Unlike traditional networks which are statically configured, SDN control applications are dynamic and are becoming more heterogeneous and complex.
	There is a great need for a framework to reason about the behavior of the various SDN applications.
	To the best of our knowledge, current network modeling frameworks were not designed to incorporate the application logic into their models, and thus can not be used to accurately model the application.
	In this paper, we suggest the possibility of leveraging the impact which control applications assert on the network information base to reason about the behavior of such applications.
	Based on that, we propose SDN-VSA, a framework that models SDN control applications as a set of affine transformations in some vector space.
	Finally, we present an analytical formulation for such framework, and discuss a use-case.
	For simplicity, we only consider the case of OpenFlow version 1.0.
\end{abstract}

\section{Introduction}\label{sec:intro}
In Software-Defined Networking (SDN), the control of the whole network is now transferred to the SDN controller which enables a set of network applications to function. It is of a great prominence to be able to analyze these applications for problems prior the actual deployment. Such problems if not fixed might take the whole network down.

A multitude of network analysis frameworks \cite{kazemian2012header,karsten2007axiomatic,dhawan2015sphinx,anderson2014netkat,nelson2015static} have been devised for analyzing traditional networks, and some of those solutions could possibly be imported into SDN. Network checking tools could be devised on top of such frameworks (\emph{e.g.,} NetPlumber \cite{kazemian2013real}), these tools can read the configurations of network devices then check these configurations for possible networking issues.

We believe that network analysis frameworks can be useful in SDN for the following reasons:
(1) SDN applications can be complex, so there is a need to reason about how would they function prior the actual deployment,
(2) controllers, in many cases, allow more than one SDN application to run simultaneously. Hence, there is a need to be able to check if those applications would create any conflict, and
(3) a \emph{what-if analysis} for any new rules could be conducted by controllers. 

We present SDN Vector Space Analysis (SDN-VSA), a framework for modeling SDN control applications. SDN-VSA is based on the idea of leveraging the impact which control applications assert on the network information base to reason about the behavior of such applications. The framework models the switching functionalities as affine transformations \cite{weisstein2017affine}. While the idea of representing data-planes as transformations is not novel and was proposed in earlier work \cite{kazemian2012header}, our framework specifically targets SDN. To realize such framework, we propose to model the control application as a composite transformation matrix operating on a network information base matrix.

In this paper, we make the following contributions:
(1) we propose an analytical framework (SDN-VSA) using vector spaces and affine transformations for modeling SDN control application,
(2) we present a formulation for such framework which can be used to analyze and reason about SDN control applications,
and (3) we discuss a use-case for the proposed framework.

The rest of this paper is organized as follows:
In \S\ref{sec:relwork}, we provide background on the topic and related work.
We discuss the need for an SDN framework in \S\ref{sec:why}.
The proposed framework is presented in \S\ref{sec:framework}.
Then, we present our formulation for such framework in \S\ref{sec:formula}.
The use-case is shown in \S\ref{sec:use-cases}
Finally \S\ref{sec:conclusion} will be our conclusion and an outline for possible foreseeable work.

\section{Related Work}\label{sec:relwork}
In this section, we survey a number of relevant frameworks which were devised for network analysis. Table \ref{tbl-survey} includes a summary of these frameworks.

Karsten \emph{et al.} \cite{karsten2007axiomatic} studied the complexity of the Internet. They found that the current Internet architecture occasionally violates the layered architecture. Thus, it is no longer appropriate to model the network as a graph. So they created an axiomatic framework intended to reason about the forwarding mechanisms in networks and to serve as a basis for verification and formal proofs of correctness of protocols and their implementations. Their formalization is based on high-level Hoare-style assertions (Hoare-logic), rather than low-level operational semantics. Further, they presented some use-cases for their framework including: TCP-over-NAT, DNS and Hierarchical Mobile IP. However, their work only considers connectivity, and is oblivious to time and cannot model losses and timeouts.

Kazemian \emph{et al.} \cite{kazemian2012header} investigated the complexity of current networks. They found that even when individual protocols function correctly, failures can arise from the complex interactions of their aggregate, requiring network administrators to be masters of detail. Thus, they came up with a protocol-agnostic framework called ``Header Space Analysis'' (HSA) based on geometric model for packet classification, with the objective of automatically finding an important class of failures, regardless of the protocols running. They also modeled the network boxes (\emph{e.g.,} switches and routers) as \emph{transfer functions} that operates on the packet headers. Their formalism allows statically checking network configurations to identify an important class of failures as Reachability Failures, Forwarding Loops and Traffic Isolation and Leakage problems. Moreover, they developed Hassel a tool that realizes HSA and was used to analyze Stanford University's backbone network. However, their model can only do static analysis of networks to detect forwarding and configuration errors, but in later work \cite{kazemian2013real} they solved that issue by developing NetPlumber.

Kazemian \emph{et al.} \cite{kazemian2013real} developed NetPlumber after they realized that the network state may change rapidly, and the network must ensure correctness. Policy checkers cannot verify in real-time because of the need to collect state and time information. They also realized that SDN provides an opportunity as it maintains a logically centralized view of the network at the controller. However, the issue of creating a fast compliance checker remains. NetPlumber is based on HSA but unlike HSA, it incrementally checks for compliance of state changes, by maintaining a dependency graph between rules (i.e. a graph of flow tables). The tool was used to detect Loops, BlackHoles and Reachability problems. They evaluated NetPlumber in (1) Google WAN, (2) Stanford Backbone, and (3) Internet2. Finally, NetPlumber (like HSA) relies on reading the state of network devices and therefore cannot model middleboxes with dynamic state, and it requires greater processing time for verifying link updates.

Dhawan \emph{et al.} \cite{dhawan2015sphinx} were concerned with the security and correct functioning of the entire SDN. They presented SPHINX to detect both known and potentially unknown attacks on network topology and data plane forwarding originating within an SDN. SPHINX leverages the novel abstraction of flow graphs for real-time detection of security threats, enable incremental validation of network updates. SPHINX analyzes specific OpenFlow control messages to learn new network behavior and metadata for both topological and forwarding state. They studied attacks on the network topology such as: ARP poisoning, and fake topology, as well as attacks on data-plane forwarding such as: Controller DoS, Network DoS, TCAM exhaustion, and switching blackholes.

Anderson \emph{et al.} \cite{anderson2014netkat} were concerned with the lack of a semantic foundation to reason precisely about networking code. They studied the network programming languages and found that the design of high-level languages for programming networks is ad-hoc, driven by the needs of applications and the capabilities of network hardware than by foundational principles. Thus, they presented NetKAT a new network programming language based on Kleene algebra a solid mathematical foundation and comes equipped with a complete equational theory. NetKAT models the network as an automaton that moves packets from one node to another along the links in its topology and hence makes use of regular expressions and the language of finite automata. As for the use-cases, they were concerned with reachability, traffic isolation, access control, and compiler correctness.

Network operators need tools to determine the impact of changes that they make, because bad updates can bring down the entire network. Nelson \emph{et al.} \cite{nelson2015static} developed Chimp a tool for static differential analysis of SDN controller programs without the need of formal methods. Chimp can be used to present the semantic or the behavioral difference between any two versions of a program. The tool was demonstrated with a NATing code written in Flowlog. Further, it was tested with a L2 learning switch, round-robin LB, and ARP cache applications. Chimp relies on Flowlog \cite{nelson2014tierless} a declarative language used for writing SDN applications that was developed by the same authors in a previous work. In Flowlog, every rule is equivalent to a first-order logic formula.

\begin{landscape}
	\begin{table}
		\caption{A comparison of some network analysis frameworks}
		\label{tbl-survey}
		\centering
		\tiny
		\begin{tabular}{p{1cm} | p{2.5cm} | p{2.5cm} | p{2.5cm} | p{2.5cm} | p{2.5cm} | p{2.5cm}}
			\hline
			
			&
			\parbox{2.5cm}{\centering Axiomatic Basis for Communication} &
			\parbox{2.5cm}{\centering Header Space Analysis} &
			\parbox{2.5cm}{\centering Real Time Network Policy Checking} &
			\parbox{2.5cm}{\centering SPHINX} &
			\parbox{2.5cm}{\centering NetKAT} &
			\parbox{2.5cm}{\centering Static Diff. Program Analysis for SDN}\\
			\hline\hline
			
			Problem
			&
			\parbox{2.5cm}{
				Internet is complex.\\
				Layering violation.
			}
			&
			\parbox{2.5cm}{
				Networks are complex.\\
				Failures can arise from complex interactions of individual protocols that function correctly.
			}
			&
			\parbox{2.5cm}{
				Networks must ensure correctness while their \emph{state} may change rapidly.\\
				Policy checkers cannot verify in real-time because they need to collect state and need time to analyze it.\\
				SDNs provide a logically centralized view but remains the need for a fast compliance checker.
			}
			&
			\parbox{2.5cm}{
				Security and correct functioning of the entire SDN.
			}
			&
			\parbox{2.5cm}{
				Design of high-level network programming languages is ad-hoc, driven by the needs of applications and capabilities of network hardware than by foundational principles.\\
				Lack of a semantic foundation to reason about networking code.
			}
			&
			\parbox{2.5cm}{
				Bad updates can bring down an entire network.
			}\\
			\hline
			
			Objective
			&
			\parbox{2.5cm}{
				Present formulation of forwarding mechanisms.\\
				Basis for verification and formal proofs of correctness of protocols and their implementations.
			}
			&
			\parbox{2.5cm}{
				Automatically find an important class of failures, regardless of the protocols running.
			}
			&
			\parbox{2.5cm}{
				Introduces a real-time policy checking tool (NetPlumber).
			}
			&
			\parbox{2.5cm}{
				Detect both known and potentially unknown attacks on network topology and data plane forwarding originating within an SDN.
			}
			&
			\parbox{2.5cm}{
				New network programming language
			}
			&
			\parbox{2.5cm}{
				Network operators therefore need tools to determine the impact of changes.
			}
			\\
			\hline
			
			Approach
			&
			\parbox{2.5cm}{
				Formalization based on high-level Hoare-style assertions (Hoare-logic), and can be used to:\\
				(1) formally analyze  network  protocols  based  on structural properties,\\
				(2) derive working prototype implementations of these protocols.
			}
			&
			\parbox{2.5cm}{
				Protocol-agnostic framework: HSA based on geometric model for packet classification.\\
				HSA allows to statically check network specs and configs to identify failures as Reachability Failures,
				Forwarding Loops and Traffic Isolation and Leakage problems.
			}
			&
			\parbox{2.5cm}{
				Based on HSA but incrementally checks for compliance of state changes, by maintaining a dependency graph between rules (a graph of flow tables).
			}
			&
			\parbox{2.5cm}{
				Analyzes specific OpenFlow control messages to learn new network behavior and metadata for both topological and forwarding state.\\
				Uses flow graphs for real-time detection of security threats, enable incremental validation of network updates.
			}
			&
			\parbox{2.5cm}{
				A new network programming language based on Kleene algebra a solid mathematical foundation and comes equipped a complete equational theory.
			}
			&
			\parbox{2.5cm}{
				Presented Chimp a tool for static differential analysis of SDN controller programs without the need of formal methods.\\
				It presents the semantic or behavioral difference between two versions of a program.\\
			}
			\\
			\hline
			
			Usecase
			&
			\parbox{2.5cm}{
				TCP over NAT\\
				DNS\\
				Hierarchical Mobile IP
			}
			&
			\parbox{2.5cm}{
				Reachability Analysis\\
				Loop Detection\\
				Slice Isolation
			}
			&
			\parbox{2.5cm}{
				Loops and BlackHoles\\
				Reachability Policies
			}
			&
			\parbox{2.5cm}{
				(1) Attacks on Network Topology: ARP Poisoning and Fake topology.\\
				(2) Attacks on Data Plane Forwarding: Controller DoS, Network DoS, TCAM exhaustion and Switch blackhole.
			}
			&
			\parbox{2.5cm}{
				Reachability\\
				Traffic isolation\\
				Access control\\
				Compiler correctness
			}
			&
			\parbox{2.5cm}{
				NAT code in Flowlog.\\
				Also: L2 Learning Switch, Round-robin LB, and ARP cache.
			}
			\\
			\hline
			
			Issues
			&
			\parbox{2.5cm}{
				Only studies connectivity.\\
				Oblivious to time and cannot model losses and timeouts.
			}
			&
			\parbox{2.5cm}{
				Only static analysis to detect forwarding and configuration errors.\\
				Offers no clues as to whether routing is efficient.
			}
			&
			\parbox{2.5cm}{
				Relies on reading the state of network devices and therefore cannot model middleboxes with dynamic state.\\
				Greater processing time for verifying link updates.
			}
			&
			\parbox{2.5cm}{
				
			}
			&
			\parbox{2.5cm}{
				
			}
			&
			\parbox{2.5cm}{
				Focus only on Flowlog (first-order logic).
			}
			\\
			\hline
			
		\end{tabular}
	\end{table}
\end{landscape}

\section{The Need for Another Framework}\label{sec:why}
Recent work in network analysis mainly handles the case of static analysis of networks in order to find important classes of failures. However, network states are dynamic and can frequently change, making those frameworks unfeasible for real-time analysis.

Some were not developed to be used during application design as they require a functioning network to monitor and analyze (\emph{e.g.,} SPHINX [12]). Those frameworks depend on monitoring the network state, which means that they can only be used after deployment.
Others do not take the application logic into their consideration, and hence fail to model dynamic stateful SDN applications, those that their current state depends on previous states such as stateful-firewalls and network address translation (NAT).

\begin{figure}
	\centering
	\includegraphics[scale=0.5]{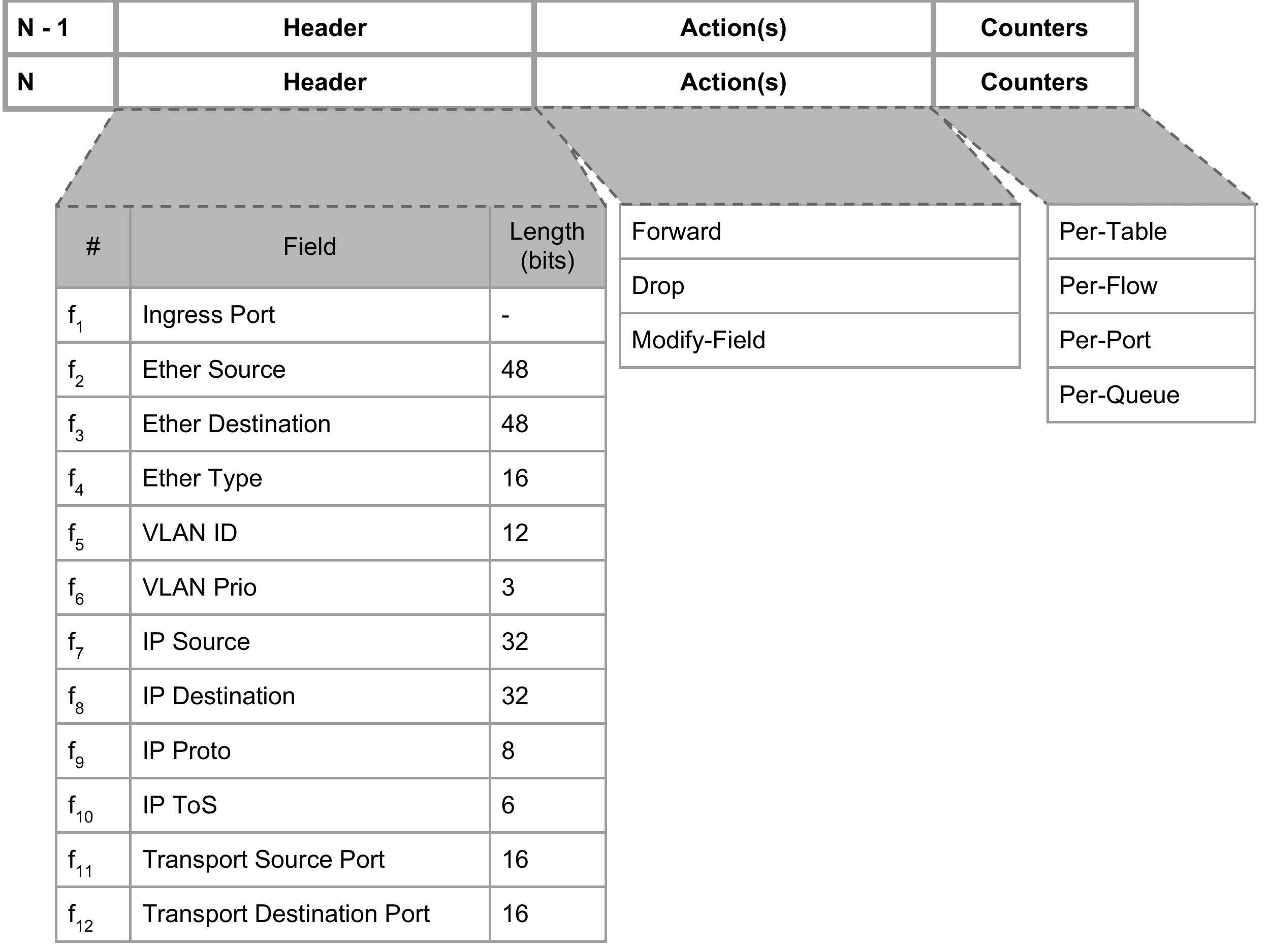}
	\caption{An OpenFlow-1.0 Table Entry; Header Fields; Actions; and Counters.}
	\label{flow-table}
\end{figure}

Moreover, we believe that one way to reason about SDN applications is through studying the impact they have on the network information base.
Therefore, on contrary to HSA \cite{kazemian2012header} which was designed to be protocol-agnostic, our model will not consider packets as first class citizens. HSA models packets as points in header space ($\mathcal{H}$) while ignoring protocol-specific meanings associated with header bits. However, in SDN OpenFlow-enabled switches we only need to consider certain fields (listed in Figure \ref{flow-table}) in the packet header such as source IP address, MAC address, and port number, and destination IP address, MAC address, and port number. The reason is that OpenFlow \cite{mckeown2008openflow} only allows matching against or modifying those fields.
As with SPHINX \cite{dhawan2015sphinx}, the control logic will be split into OpenFlow \cite{mckeown2008openflow} primitive control messages.
Each control message should be studied carefully and represented as a transform that would change the current network state.

\section{The SDN-VSA Framework}\label{sec:framework}
The proposed framework (SDN-VSA) assumes that SDN control applications are \emph{deterministic} applications (\emph{i.e.,} they produce the same output for a certain input). Hence, the framework can leverage the impact (\emph{i.e.,} via the OpenFlow messages they send to the switches) that applications have on the network information base to reason about such applications.
In this section, we make some important definitions, and we present the main axioms of the proposed framework which are required in order to model SDN control applications.

\subsection{Definitions}\label{defs}
\begin{definition}
	\textbf{Control Application:}
	is a set of instructions running at the control-plane which has the ability to control the switching functionalities of connected data-planes and collect information from them via messages of a standard protocol (\emph{e.g.,} OpenFlow).
\end{definition}
\begin{definition}
	\textbf{Action:}
	is an instructions associated with a header, installed by the control application and executed by the data-plane when a flow arrives matching that header. In case of OpenFlow, actions are installed by FLOW\_MOD messages.
\end{definition}
\begin{definition}
	\textbf{Network Information Base (NIB):}
	is the aggregate of topology and/or state information that allows a control application to function properly.
\end{definition}

\subsection{Axioms}\label{axioms}
The following axioms represent a basis for the SDN-VSA framework:

\begin{axiom}
	Any OpenFlow message can be modeled as a linear map in some vector space.
	\label{axiom1}
\end{axiom}
Let $M$ be the set of all OpenFlow messages, and $V$ is some vector space over a field $F$, then:
\begin{center}
	$\forall m \in \mathcal{M}$, $\exists V^{n}$ $\wedge$ $\exists T_{m}: V^{n}\rightarrow V^{n}$ $\mid$\\
	$T_{m}(p_{1} + p_{2}) = T_{m}(p_{1}) + T_{m}(p_{2})$ $\dots$ $p_{1}, p_{2} \in V$\\
	$\wedge$
	$T_{m}(c.p) = c . T_{m}(p)$ $\dots$ $c \in F, p \in V$
\end{center}

\begin{axiom}
	Any OpenFlow action can be modeled as a linear map in some vector space (n-dimensional space).
	\label{axiom2}
\end{axiom}
Let $\mathcal{A}$ be the set of all OpenFlow actions, and $V$ is some vector space over a field $F$, then:
\begin{center}
	$\forall a \in \mathcal{A}$, $\exists V^{n}$ $\wedge$ $\exists T_{a}: V^{n}\rightarrow V^{n}$ $\mid$\\
	$T_{a}(p_{1} + p_{2}) = T_{a}(p_{1}) + T_{a}(p_{2})$ $\dots$ $p_{1}, p_{2} \in V$\\
	$\wedge$
	$T_{a}(c.p) = c . T_{a}(p)$ $\dots$ $c \in F, p \in V$
\end{center}

\subsection{The Model}\label{model}

From the definitions presented in \S\ref{defs} and the Axioms \ref{axiom1} and \ref{axiom2} proposed in \S\ref{axioms}, we can deduce the following corollaries:

\begin{corollary}
	A control application can be modeled as a composite transformation matrix that operates on a network information base matrix.
	\label{coro1}
\end{corollary}

\begin{equation}
\begin{bmatrix}
t^{'}_{1}\\
t^{'}_{2}\\
\vdots \\
t^{'}_{n-1} \\
1
\end{bmatrix}
=
\begin{bmatrix}
c_{1, 1} & c_{1, 2} & \cdots & c_{1, n}\\
c_{2, 1} & c_{2, 2} & \cdots & c_{2, n}\\
\vdots  & \vdots  & \ddots & \vdots \\
c_{n-1, 1} & c_{n-1, 2} & \cdots & c_{n-1, n} \\
0 & 0 & \cdots & 1
\end{bmatrix}
\times
\begin{bmatrix}
t_{1}\\
t_{2}\\
\vdots \\
t_{n-1} \\
1
\end{bmatrix}
\end{equation}

\begin{corollary}[The Congruence Principle]
	Any two control applications are congruent in a certain vector space if-and-only-if their composite transformation matrices in that vector space are equal.
	\label{coro2}
\end{corollary}

\section{The Formulation}\label{sec:formula}
In this section, we present a formulation for the SDN-VSA framework, other formulations might also exist. Depending on the control applications, the NIB might include more network information (\emph{i.e.,} application-specific or non-OpenFlow) other than just the flow tables. However, in this paper we only formulate flows tables. More specifically, we define $\mathcal{T}$-space $(\mathcal{T}, +, \times)$, a vector space over a Galois field \cite{lidl1997finite} of two elements (GF(2)) (for simplicity) where control messages can be represented as linear maps as postulated by Axiom \ref{axiom1}. Then, we formulate $\mathcal{R}$-space as postulated by Axiom \ref{axiom2}.

\subsection{Assumptions}
For simplicity, we make the following assumptions: 
\begin{itemize}
	\item OpenFlow counters can be: (1) per-table, (2) per-flow, (3) per-port, and (4) per-queue counters. However, in this formulation, we only consider per-flow counters (we assume a single per-flow counter).
	\item We assume that a switch can have a single flow-table. Hence, a switch can be seen equivalent to a flow table.
	\item We are not considering queues and Quality of Service (QoS) in this formulation.
	\item We assume the network topology to be static \emph{i.e.,} no switches can be removed or added.
\end{itemize}

\subsection{Flow Tables Space ($\mathcal{T}$-space)}
Let $\mathcal{T}$ be the set of all flow tables.
A flow table $t \in \mathcal{T}$ can be modeled as a set of tuples (\emph{i.e.,} ordered-pairs) of rules ($r$) and counters ($c$) (see Figure \ref{flow-table}). Therefore, we define $t$ as:
\begin{equation}
t = \{(r_{1}, c_{1}), \dots, (r_{n}, c_{n})\} \dots \forall t \in \mathcal{T}, \forall r \in \mathcal{R}
\end{equation}

Let $t, t_{1}, t_{2}, \Phi \in \mathcal{T}$. Let $+, \times$ be two operations (vector addition and scalar multiplication, respectively) such that $\forall t_{1},t_{2} \in \mathcal{T}$, $t_{1} + t_{2} \in \mathcal{T}$ and $\forall a \in GF(2)$, $a \times t \in \mathcal{T}$. We define the vector addition using basic \emph{set theory} as follows:
\begin{eqnarray}
t_{1} + t_{2} = t_{1} \cup t_{2} \dots \forall t_{1}, t_{2} \in \mathcal{T} \\
t + \Phi = t \dots \forall t, \Phi \in \mathcal{T} 
\end{eqnarray}
We define the scalar multiplication as follows:
\begin{eqnarray}
0 \times t = \Phi \dots \forall t, \Phi \in \mathcal{T} \\
1 \times t = t \dots \forall t \in \mathcal{T}
\end{eqnarray}

Moreover, for every flow table $t \in \mathcal{T}$ with a number ($n$) of flow rules ($\in \mathcal{R}$), we can define another flow table $(-t) \in \mathcal{T}$ with an equal number ($n$) of flow rules, such that every flow rule ($r \in \mathcal{R}$) in $t$ has an \emph{additive inverse} ($-r \in \mathcal{R}$) in $(-t)$.

Since, the following conditions are satisfied \cite{weisstein2017vector}:
\begin{eqnarray}
(t_{1} + t_{2}) + t_{3} = t_{1} + (t_{2} + t_{3}) \dots \forall t_{1}, t_{2}, t_{3} \in \mathcal{T} \\
t_{1} + t_{2} = t_{2} + t_{1} \dots \forall t_{1}, t_{2} \in \mathcal{T} \\
\exists \Phi \in \mathcal{T} \mid t + \Phi = t \dots \forall t \in \mathcal{T} \\
\exists -t \in \mathcal{T} \mid t + (-t) = \Phi \dots \forall t \in \mathcal{T} \\
a \times (b \times t) = (a \times b) \times t \dots \forall a, b \in GF(2), t \in \mathcal{T} \\
\exists 1 \in GF(2) \mid 1 \times t = t \dots \forall t \in \mathcal{T} \\
a \times (t_{1} + t_{2}) = a \times t_{1} + a \times t_{2} \dots \forall a \in GF(2), t_{1}, t_{2} \in \mathcal{T} \\
(a + b) \times t = a \times t + b \times t \dots \forall a, b \in GF(2), t \in \mathcal{T} 
\end{eqnarray}

Therefore, $\mathcal{T}$-space $(\mathcal{T}, +, \times)$ is a vector space over $GF(2)$ a Galois field of two elements. $\blacksquare$ 

\subsection{Flow Rules Space ($\mathcal{R}$-space)}\label{subsec:r-space}
Let $\mathcal{R}$ be the set of all flow rules. A flow rule $r \in \mathcal{R}$ can be modeled as a vector (tuple) of: (1) a header $h \in \mathcal{H}^{n}$ (to match against) where $n$ is the number of OpenFlow header fields, (2) an output port $p^{-}$, (3) a time-to-live $\tau$ (we assume one type of TTL), and (4) a set of actions $\alpha \in \mathcal{A}$ (a composite transformation matrix).
\begin{equation}
r = <h, p^{-}, \tau, \alpha> \dots \forall r \in \mathcal{R}
\end{equation}

In HSA \cite{kazemian2012header}, the $\mathcal{H}$-space is protocol-agnostic. Hence, it views a packet header as a sequence of ones and zeros in an $\ell$-dimensional space ($\{0, 1\}^\ell$) ($\ell$ is the header length in bits).
We model \footnote{In this paper, we are not going to prove that the $\mathcal{R}$-space satisfies the conditions of a vector space, as the vector spaces of the individual components comprising a flow rule are well-known.} a header $h \in \mathcal{H}^{n}$ in a similar way, but we only consider the case of OpenFlow. In particular, we represent a header $h$ as a tuple of the $n$ OpenFlow's fixed-length header fields (see Figure \ref{flow-table}):
\begin{equation}
h = <f_{1}, f_{2}, ....., f_{n}>  \dots \forall h \in \mathcal{H}^{n}, \forall f_{i} \in \mathcal{H}
\end{equation}

An action $\alpha \in \mathcal{A}$ is a function in the flow header $h \in \mathcal{H}^{n}$, hence it can be modeled as follows:
\begin{equation}
\alpha(h)
\times
\begin{bmatrix}
h \\
p^{-} \\
\tau \\
1
\end{bmatrix}
\end{equation}

Finally, for every flow rule $r \in \mathcal{R}$ with a set of actions $\alpha \in \mathcal{A}$, we can define another flow rule $(-r) \in \mathcal{R}$ with the same flow header $h \in \mathcal{H}^n$ and the same output port $p^-$, such that $(-r)$ has a set of actions $\alpha^{-1} \in \mathcal{A}$. Let $I$ be the identity matrix, then:
\begin{equation}
\alpha(h) \times \alpha^{-1}(h) = I
\end{equation}
We call $(-r)$ the \emph{additive inverse} of $r$.
\begin{equation}
\exists -r \in \mathcal{R} \mid r + (-r) = \phi \dots \forall r, \phi \in \mathcal{R} \\
\end{equation}

\subsection{Example: FLOW\_MOD}
In OpenFlow, a FLOW\_MOD message can be used to add, modify, or delete a flow rule from a flow table.
In case of add or modify commands, an action list needs to be specified. The following actions are supported by OpenFlow:

\subsubsection{Forward}
An OFPAT\_OUTPUT action is responsible for outputting any matched flow to a specific switch port \emph{i.e.,} forwarding.
A forward action ($\alpha_{f}(h)$) can be modeled as a translation of the output port ($p^{-}$) by $\delta$ as follows:
\begin{equation}
\alpha_{f}(h)
=
\begin{bmatrix}
1 & 0 & 0 & 0 \\
0 & 1 & 0 & \delta \\
0 & 0 & 1 & 0 \\
0 & 0 & 0 & 1
\end{bmatrix}
\end{equation}

\subsubsection{Drop}
In OpenFlow, a packet belonging to a matched flow is dropped when the flow rule has an empty action list.
Therefore, a flow dropping action ($\alpha_{d}(h)$) can be modeled as a zero-scaling transformation as follows:
\begin{equation}
\alpha_{d}(h)
=
\begin{bmatrix}
0 & 0 & 0 & 0 \\
0 & 0 & 0 & 0 \\
0 & 0 & 0 & 0 \\
0 & 0 & 0 & 1
\end{bmatrix}
\end{equation}

\subsubsection{Modify-Field}
In OpenFlow, certain header fields ($\in \mathcal{H}^{n}$) can be modified (see Figure \ref{flow-table}) by a modify-field action.
Hence, a modify-field action ($\alpha_{m}(h)$) can be modeled as a translation of the flow header $h$ by $\delta \in \mathcal{H}^{n}$ as follows:
\begin{equation}
\alpha_{m}(h)
=
\begin{bmatrix}
1 & 0 & 0 & \delta \\
0 & 1 & 0 & 0 \\
0 & 0 & 1 & 0 \\
0 & 0 & 0 & 1
\end{bmatrix}
\end{equation}

\section{Use-cases}\label{sec:use-cases}


\subsection{SDN Service Chaining}
With the increasing popularity of SDN as an enabler technology for Network Function Virtualization (NFV) in the Cloud, network service providers tend to implement their middle-boxed network services as Virtualized Network Functions (VNF) \cite{bari2015orchestrating}. Such services include but are not limited to load-balancers, firewalls and Intrusion Detection Systems (IDS) \cite{joseph2008policy}. The process of chaining those services is known as \emph{service chaining}.

Oftentimes, network service providers - for many reasons - tend to \emph{steer} different flows across different sets of middle-boxes (\emph{i.e.,} services) \cite{zhang2013steering}.
Recent research is concerned with the issues of steering packets through different middle-boxes \cite{chukwu2017one}.
Moreover, in some cases, the order at which the flows steer the network services matters.
For example, in the simple case of two services (show in Figure \ref{servicechain}): a load-balancer and an IDS, it is often more appropriate to steer incoming flows through the IDS first then the load-balancer in order to reduce the latency (the NIB used in this paper does not count-in latencies) or prevent any malicious flows from reaching the load-balanced servers.
In such case, we believe that the proposed framework can be used to reason about flow steering in SDN-enabled network services.

As network services are also control applications, then based on Corollary \ref{coro1} (see \S\ref{model}) of the proposed framework, they can be modeled individually as composite transformation matrices.
Additionally, we can simply deduce the following Corollary:

\begin{corollary}
	Any chain of services can be modeled as a composite transformation matrix of its set of services that operates on a network information base matrix in some vector space.
	\label{coro3}
\end{corollary}

We can also deduce from Corollary \ref{coro2} (see \S\ref{model}) ``The Congruence Principle'' that any two chains of services are congruent in a certain vector space if-and-only-if their composite transformation matrices in that vector space are equal.

Moreover, based on the fact that matrix multiplication is non-commutative, different orders of same services can yield different composite transformation matrices. However, as translation is isomorphic in $\mathcal{T}$-space $(\mathcal{T}, +, \times)$, order does not matter in any service chain that only contains translations.

For example, assume the following two service chains (shown in Figure \ref{servicechain}): (1) IDS $\rightarrow$ LB (shown in blue in Figure \ref{servicechain}), and (2) LB $\rightarrow$ IDS (shown in red in Figure \ref{servicechain}).
Let $A_{f}$ be a forwarding action, $A_d$ be a drop action, and $A_{m,f}$ be a composite action of modify then forward actions.
Let $N(h)$ be a function which given a certain header $h$, returns the total number of flows $f_i$ having the same source field as $h$.
And let $\nu$ be the anomaly detection threshold. 
\begin{equation}
N(h) = \sum_{i=1} [SRC(f_{i}) = SRC(h)] 
\end{equation}
Let $L(s)$ be a function that given a certain server's IP address $s$, returns the total number of flows $f_i$ currently handled by that server.
\begin{equation}
L(s) = \sum_{i=1} [DEST(f_{i}) = s] 
\end{equation}

In the first case (IDS $\rightarrow$ LB), the control transformation matrices ($\mathcal{X}_{IDS}$ and $\mathcal{X}_{LB}$) would be:
\begin{equation}
\mathcal{X}_{IDS}(t, h)
=
\begin{cases}
t + \{(<h, p_{LB}, \tau, A_f>, 0)\}, & \mbox{if }  N(h) \le \nu \\
t + \{(<h, p_{LB}, \tau, A_d>, 0)\}, & \mbox{otherwise}
\end{cases}
\end{equation}

\begin{equation}
\mathcal{X}_{LB}(t, h)
=
\begin{cases}
t + \{(<h, p_{s_{1}}, \tau, A_{m,f}>, 0)\}, & \mbox{if } L(s_{1}) \le L(s_{2}) \\
t + \{(<h, p_{s_{2}}, \tau, A_{m,f}>, 0)\}, & \mbox{otherwise}	
\end{cases}
\end{equation}

\begin{equation}
\begin{gathered}
\mathcal{X}_{IDS \rightarrow LB}
=
\begin{bmatrix}
1 & 0 & 0 \\
0 & 1 & \mathcal{X}_{IDS} \\
0 & 0 & 1
\end{bmatrix}
\times
\begin{bmatrix}
1 & 0 & \mathcal{X}_{LB} \\
0 & 1 & 0 \\
0 & 0 & 1
\end{bmatrix}
=\\
\begin{bmatrix}
1 & 0 & \mathcal{X}_{LB} \\
0 & 1 & \mathcal{X}_{IDS} \\
0 & 0 & 1
\end{bmatrix}
\end{gathered}
\label{eqn:IDS-LB}
\end{equation}
Where $p_{LB}$ is the load-balancer's port. $p_{s_{1}}$, and $p_{s_{2}}$ are the first and second server's ports, respectively.

In the second case (LB $\rightarrow$ IDS), the control transformation matrices ($\mathcal{Y}_{IDS}$ and $\mathcal{Y}_{LB}$) would be:
\begin{equation}
\mathcal{Y}^{'}_{IDS}(t, h) = t + \{(<h, p_{LB}, \tau, A_f>, 0)\}
\end{equation}

\begin{equation}
\mathcal{Y}^{'}_{LB}(t, h)
=
\begin{cases}
t + \{(<h, p_{IDS}, \tau, A_{m,f}>, 0)\}, & \mbox{if } L(s_{1}) \le L(s_{2}) \\
t + \{(<h, p_{IDS}, \tau, A_{m,f}>, 0)\}, & \mbox{otherwise}
\end{cases}
\end{equation}

\begin{equation}
\mathcal{Y}^{''}_{IDS}(t, h)
=
\begin{cases}
t + \{(<h, p_{LB}, \tau, A_f>, 0)\}, & \mbox{if }  N(h) \le \nu \\
t + \{(<h, p_{LB}, \tau, A_d>, 0)\}, & \mbox{otherwise}
\end{cases}
\end{equation}

\begin{equation}
\mathcal{Y}^{''}_{LB}(t, h) = t + \{(<h, p_{DEST(h)}, \tau, A_f>, 0)\}
\end{equation}
Where $p_{IDS}$ is the IDS's port. $p_{DEST_{h}}$ is the port of server with an IP address $DEST(h)$.

\begin{equation}
\begin{gathered}
\mathcal{Y}_{LB \rightarrow IDS}
=
\begin{bmatrix}
1 & 0 & 0 \\
0 & 1 & \mathcal{Y}{''}_{LB} \\
0 & 0 & 1
\end{bmatrix}
\times
\begin{bmatrix}
1 & 0 & \mathcal{Y}{''}_{IDS} \\
0 & 1 & 0 \\
0 & 0 & 1
\end{bmatrix}
\times \\
\begin{bmatrix}
1 & 0 & 0 \\
0 & 1 & \mathcal{Y}{'}_{LB} \\
0 & 0 & 1
\end{bmatrix}
\times
\begin{bmatrix}
1 & 0 & \mathcal{Y}{'}_{IDS} \\
0 & 1 & 0 \\
0 & 0 & 1
\end{bmatrix}
=\\
\begin{bmatrix}
1 & 0 &  \mathcal{Y}{'}_{IDS} + \mathcal{Y}{''}_{IDS}\\
0 & 1 &  \mathcal{Y}{'}_{LB} + \mathcal{Y}{''}_{LB}\\
0 & 0 & 1
\end{bmatrix}
\end{gathered}
\label{eqn:LB-IDS}
\end{equation}

From (\ref{eqn:IDS-LB}) and (\ref{eqn:LB-IDS}), the composite transformation matrices for the two service chains are not equal $\mathcal{X}_{IDS \rightarrow LB} \neq \mathcal{Y}_{LB \rightarrow IDS}$.
Therefore, the two service chains: IDS $\rightarrow$ LB, and LB $\rightarrow$ IDS are not congruent \emph{i.e.,} they have different impact on the network information base.

\usetikzlibrary{chains}
\usetikzlibrary{decorations.text}
\usetikzlibrary{decorations.pathmorphing}
\usetikzlibrary{scopes}

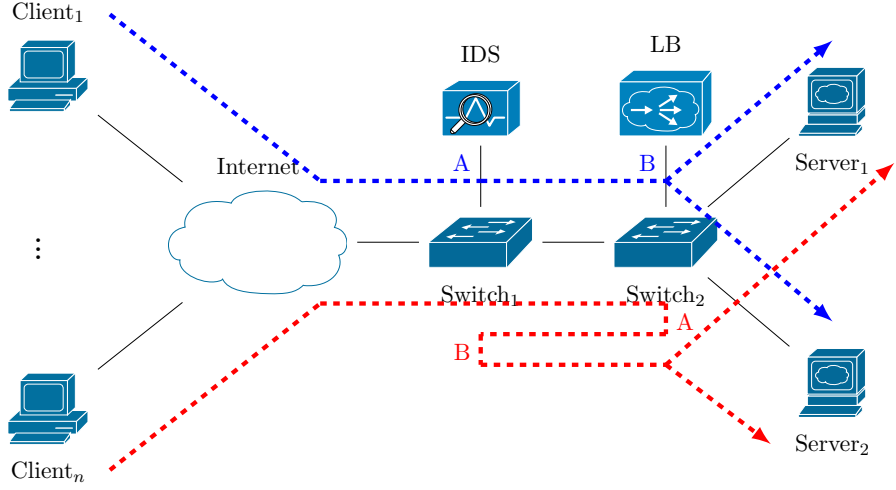
\begin{figure}[!t]
	\centering
	\resizebox{.95\textwidth}{!}{
		\begin{tikzpicture}[
		start chain=going right,
		diagram item/.style={
			on chain,
			join
		}
		]
		
		\node[label=above:Internet, diagram item] (inet) {\cloud};
		{ [start branch=A going right]
			\node[label=above:Client$_{1}$, start branch=1 going above left, diagram item] (client1) {\client};
			\node[label=below:Client$_{n}$, start branch=2 going below left, diagram item] (client2) {\client};
			\node[start branch=3 going left, on chain] (dots) {\textbf{\vdots}$\qquad$};
			
			\node[label=below:Switch$_{1}$, diagram item] (switch1) {\switch};
			\node[label=above:IDS, start branch=4 going above, diagram item] (ids) {\ids};
			
			\node[label=below:Switch$_{2}$, diagram item] (switch2) {\switch};
			\node[label=above:LB, start branch=5 going above, diagram item] (lb) {\lb};
			
			\node[label=below:Server$_{1}$, start branch=6 going above right, diagram item] (server1) {\server};
			\node[label=below:Server$_{2}$, start branch=7 going below right, diagram item] (server2) {\server};
		}
		\draw [dashed, blue, line width=0.7mm] ($(client1) + (1,1)$) -- ($(inet) + (1,1)$);
		\draw [dashed, blue, line width=0.7mm] ($(inet) + (1,1)$) -- ($(switch1) + (0,1)$) node[above left] {A};
		\draw [dashed, blue, line width=0.7mm] ($(switch1) + (0,1)$) -- ($(switch2) + (0,1)$) node[above left] {B};
		\draw [dashed, blue, -latex,line width=0.7mm] ($(switch2) + (0,1)$) -- ($(server1) + (0,1)$);
		\draw [dashed, blue, -latex,line width=0.7mm] ($(switch2) + (0,1)$) -- ($(server2) + (0,1)$);
		
		\draw [dashed, red, line width=0.7mm] ($(client2) + (1,-1)$) -- ($(inet) + (1,-1)$);
		\draw [dashed, red, line width=0.7mm] ($(inet) + (1,-1)$) -- ($(switch2) + (0,-1)$) node[below right] {A};
		\draw [dashed, red, line width=0.7mm] ($(switch2) + (0,-1)$) -- ($(switch2) + (0,-1.5)$);
		\draw [dashed, red, line width=0.7mm] ($(switch2) + (0,-1.5)$) -- ($(switch1) + (0,-1.5)$) node[below left] {B};
		\draw [dashed, red, line width=0.7mm] ($(switch1) + (0,-1.5)$) -- ($(switch1) + (0,-2)$);
		\draw [dashed, red, line width=0.7mm] ($(switch1) + (0,-2)$) -- ($(switch2) + (0,-2)$);
		\draw [dashed, red, -latex,line width=0.7mm] ($(switch2) + (0,-2)$) -- ($(server2) + (-1,-1)$);
		\draw [dashed, red, -latex,line width=0.7mm] ($(switch2) + (0,-2)$) -- ($(server1) + (1,-1)$);
		\end{tikzpicture}
	}
	\caption{Two Chains of Network Services.}
	\label{servicechain}
\end{figure}

\subsection{Discussion on Uses-Cases}
In this paper, we only presented the use-case of analyzing flow steering in network service chains. However, the proposed framework could be used for other uses-cases.
For example, we believe that the SDN-VSA framework can be used in the detection of forwarding loops. A forward loop is created when a packet (with an unchanged header) returns to a port it previously visited \cite{kazemian2012header}. Using SDN-VSA, a forwarding loop could be detected by analyzing the composite transformation matrix. In particular, by scanning the flow tables and looking for flow rules that have additive inverses (see \S\ref{subsec:r-space}).

\section{Conclusion and Future Work}\label{sec:conclusion}
In this paper, we proposed SDN-VSA, a framework that uses affine transformations and vector spaces to model SDN control applications. Then, we presented a formulation for the framework that can be used in the analysis of SDN control applications. Finally, we showed a uses-case for the SDN-VSA framework.
In the future, we plan to explore more use-cases for such framework. Nonetheless, we plan to explore other formulations for the SDN-VSA framework using different vector spaces which can support more complex OpenFlow functionalities as QoS.

\section*{Acknowledgment}
The second author acknowledges support from the Natural Sciences and Engineering Research Council of Canada (NSERC) through the NSERC Discovery Grant program. The authors would like to thank Jason Jaskolka for his invaluable remarks on an earlier draft of the paper.

\bibliographystyle{abbrv}
\bibliography{references}

\begin{thebibliography}{10}

\bibitem{anderson2014netkat}
C.~J. Anderson, N.~Foster, A.~Guha, J.-B. Jeannin, D.~Kozen, C.~Schlesinger,
  and D.~Walker.
\newblock Netkat: Semantic foundations for networks.
\newblock {\em ACM SIGPLAN Notices}, 49(1):113--126, 2014.

\bibitem{bari2015orchestrating}
M.~F. Bari, S.~R. Chowdhury, R.~Ahmed, and R.~Boutaba.
\newblock On orchestrating virtual network functions.
\newblock In {\em Network and Service Management (CNSM), 2015 11th
  International Conference on}, pages 50--56. IEEE, 2015.

\bibitem{chukwu2017one}
J.~C. Chukwu and A.~Matrawy.
\newblock One pass packet steering (opps) for stateless policy chains in
  multi-subscriber sdn.
\newblock In {\em 2017 IEEE Conference on Computer Communications Workshops
  (INFOCOM WKSHPS)}, May 2017.

\bibitem{dhawan2015sphinx}
M.~Dhawan, R.~Poddar, K.~Mahajan, and V.~Mann.
\newblock Sphinx: Detecting security attacks in software-defined networks.
\newblock In {\em Proceedings of the 2015 Network and Distributed System
  Security (NDSS) Symposium}, 2015.

\bibitem{joseph2008policy}
D.~A. Joseph, A.~Tavakoli, and I.~Stoica.
\newblock A policy-aware switching layer for data centers.
\newblock In {\em ACM SIGCOMM Computer Communication Review}, volume~38, pages
  51--62. ACM, 2008.

\bibitem{karsten2007axiomatic}
M.~Karsten, S.~Keshav, S.~Prasad, and M.~Beg.
\newblock An axiomatic basis for communication.
\newblock In {\em ACM SIGCOMM Computer Communication Review}, volume~37, pages
  217--228. ACM, 2007.

\bibitem{kazemian2013real}
P.~Kazemian, M.~Chan, H.~Zeng, G.~Varghese, N.~McKeown, and S.~Whyte.
\newblock Real time network policy checking using header space analysis.
\newblock In {\em NSDI}, pages 99--111, 2013.

\bibitem{kazemian2012header}
P.~Kazemian, G.~Varghese, and N.~McKeown.
\newblock Header space analysis: Static checking for networks.
\newblock In {\em NSDI}, pages 113--126, 2012.

\bibitem{lidl1997finite}
R.~Lidl and H.~Niederreiter.
\newblock {\em Finite fields}, volume~20.
\newblock Cambridge university press, 1997.

\bibitem{mckeown2008openflow}
N.~McKeown, T.~Anderson, H.~Balakrishnan, G.~Parulkar, L.~Peterson, J.~Rexford,
  S.~Shenker, and J.~Turner.
\newblock Openflow: enabling innovation in campus networks.
\newblock {\em ACM SIGCOMM Computer Communication Review}, 38(2):69--74, 2008.

\bibitem{nelson2015static}
T.~Nelson, A.~D. Ferguson, and S.~Krishnamurthi.
\newblock Static differential program analysis for software-defined networks.
\newblock In {\em FM 2015: Formal Methods}, pages 395--413. Springer, 2015.

\bibitem{nelson2014tierless}
T.~Nelson, A.~D. Ferguson, M.~J. Scheer, and S.~Krishnamurthi.
\newblock Tierless programming and reasoning for software-defined networks.
\newblock {\em NSDI, Apr}, 2014.

\bibitem{weisstein2017affine}
E.~W. Weisstein.
\newblock Affine transformation.
\newblock 2017.
\newblock \url{http://mathworld.wolfram.com/AffineTransformation.html}.

\bibitem{weisstein2017vector}
E.~W. Weisstein.
\newblock Vector space.
\newblock 2017.
\newblock \url{http://mathworld.wolfram.com/VectorSpace.html}.

\bibitem{zhang2013steering}
Y.~Zhang, N.~Beheshti, L.~Beliveau, G.~Lefebvre, R.~Manghirmalani, R.~Mishra,
  R.~Patneyt, M.~Shirazipour, R.~Subrahmaniam, C.~Truchan, et~al.
\newblock Steering: A software-defined networking for inline service chaining.
\newblock In {\em Network Protocols (ICNP), 2013 21st IEEE International
  Conference on}, pages 1--10. IEEE, 2013.

\end{thebibliography}

\end{document}